\documentclass [] {aa} 
\usepackage{epsfig, graphicx}
\usepackage{natbib}

\begin{document}

%\thesaurus{06(08.01.1; 08.03.02; 08.16.2)}

%\titlerunning{} 
\title{Galactic Evolution of Nitrogen.}

%\subtitle{}

\author{G.~Israelian\inst{1}, A.~Ecuvillon\inst{1}, R.~Rebolo\inst{1,2}, 
R.~Garc\'{\i}a-L\'opez\inst{1,3}, P.~Bonifacio\inst{4} \and 
P.~Molaro\inst{4}}

\offprints{ \email{gil@ll.iac.es}}

\institute{Instituto de Astrof\'{\i}sica de Canarias, E--38200 La Laguna, Tenerife, Spain 
\and Consejo Superior de Investigaciones Cient\'\i fcas, Spain
\and Departamento de Astrof\'{\i}sica, Universidad de La Laguna, Av. Astrof\'{\i}sico 
Francisco S\'anchez s/n, E-38206 La Laguna, Tenerife, Spain 
\and Osservatorio Astronomico di Trieste, via G. B. Tiepolo 11, 34131
Trieste, Italy}

\date{Received / Accepted } 

\titlerunning{Galactic Evolution of Nitrogen.} 
\authorrunning{G. Israelian et al.}
%--------------------------------------------------------------------------

\abstract{
We present detailed spectroscopic analysis of nitrogen abundances in 31 unevolved metal-poor stars 
analysed by spectral synthesis of the near-UV NH band at 3360 \AA\ observed at high resolution with 
various telescopes. We found that [N/Fe] scales with that of iron in the metallicity 
range $-3.1 <$[Fe/H]$<0$ with the slope 0.01$\pm$0.02. Furthermore, we derive uniform and accurate 
(N/O) ratios using oxygen abundances from near-UV OH lines obtained in our previous studies. 
We find that a primary component of nitrogen is required to explain the observations. 
The NH lines are discovered in the VLT/UVES spectra of the very metal-poor subdwarfs G64-12 and LP815-43
indicating that these stars are N rich. The results are compared with theoretical models and observations
of extragalactic HII regions and Damped Ly$\alpha$ systems. This is the first direct comparison of the (N/O) 
ratios in these objects with those in Galactic stars. 
\keywords{stars: abundances -- 
          stars: nucleosynthesis -- 
	  Galaxy: evolution
	  Galaxy: halo
	  }}
\maketitle
\section{Introduction}
\label{Intro}

Elemental abundance studies are important tracers of galaxy formation and evolution. The yields and 
origin of some important elements, such as nitrogen, are still a matter of serious debate. The stable isotope 
\element[][14]{N} is synthesized from $^{12}$C and $^{16}$O through the CNO cycles in a hydrogen-burning layer of a star.
Knowledge of the origin of nitrogen is very important since, for example, this element can be used to derive the primordial 
helium abundance (Pagel \& Kazlauskas 1992). We call nitrogen ``primary'' when it is produced solely from the original 
hydrogen and helium in a star, either directly or through successive stages of nucleosynthesis.  When the effective 
yield of nitrogen depends on the previous enrichment of C and O, the nitrogen is considered as ``secondary''. 
Which stars produce secondary nitrogen is fairly well understood: non-zero metal-rich intermediate mass stars in the H-shell
during the AGB phase (van den Hoek \& Groenewegen 1997). Freshly produced nitrogen may be brought to the surface by the
third dredge-up and released in the interstellar medium by the stellar wind. Instead there are at least two possible sources
of primary nitrogen:

\begin{enumerate}
\item intermediate mass which either undergo Hot Bottom Burning (Marigo 2001) or rotate (Meynet \& Maeder 2002);
\item metal-poor fast rotating massive stars where primary N is produced during the He-burning 
phase by rotational diffusion of $^{12}$C into the He-burning layer (Meynet \& Maeder 2002). 
\end{enumerate}

The relative weight of secondary and primary components depends on the interplay between secondary enrichment caused by 
dredge up episodes and the primary contribution given by the CNO cycle during the envelope burning. It is well known that 
oxygen and perhaps also carbon are produced by massive stars (Maeder 1992).  Several observations of the Galactic halo 
stars provided some support on the primary origin of N at low metallicities (Edmunds \& Pagel 1978; Pagel \& Edmunds 1981;
Bessel \& Norris 1982; Tomkin \& Lambert 1984; Laird 1985; Carbon et al.\ 1987). However, it is commonly accepted that 
the main observational evidence of primary nitrogen at low metallicities is provided from abundance studies of O and N 
in H\,II regions of nearby metal-poor galaxies. Sargent \& Searle (1970) proposed that the most metal-poor
galaxies are young systems experiencing their first star formation phase. At metallicities lower than 
log(O/H)$\leq -$4.0 nitrogen is mostly primary with log(N/O)$\simeq -$1.5 (see Henry et al. 2000, and references therein). 
This fact has led Thuan, Izotov \& Lipovetsky (1995) to propose that N in these galaxies 
has been produced by massive stars as a primary element, and that therefore these stellar systems are young. 
However, recent spectrophotometric studies (e.g. Schulte-Ladbeck et al.\ 2001) of some metal-poor blue 
compact dwarf galaxies (BCGs) revealed old stellar populations. The cases of I\,Zw 18 and SBS\,0335--052 are still a 
matter of debate (i.e. Hunt, Thuan \& Izotov 2003). We also note that Henry et al. (2000) have proposed a different 
explanation for the low metallicity (N/O) plateau. These authors suggested that a very low star formation rate with 
massive stars producing O and intermediate mass stars producing primary N with a time-lag of $\sim$ 250 Myr may 
produce a constant (N/O) as well. In fact, using the yields of N (van den Hoek \& Groenewegen 1997) and 
O (Maeder 1992) for a standard IMF, one obtains log(N/O)$\sim -$1.5 at log(O/H)$< -$4 (Henry et al. 2000). 
Several authors have concluded that the contribution by massive stars to nitrogen is generally very low and the 
dominant primary N contributor is intermediate mass stars  (Henry et al.\ 2000; Liang et al.\ 2001). 
However, others support the origin of primary nitrogen in massive stars (e.g. Thuan et al.\ 1995; Izotov et al.\ 2001)
because of the low scatter of (N/O) ratios in galaxies observed at different stages of 
their evolution, which would imply no time delay between injection of nitrogen and oxygen. The scatter of
(N/O) is large in high redshift low metallicity damped Ly $\alpha$ systems (DLAs) (Centuri\'on et al.\ 1998; Lu, Sargent \& 
Barlow 1998; Pettini et al.\ 2002). 
However the sample can be split into two groups, each one with a relatively small dispersion (Centuri\'on et al.\ 2003). The
group with lower (N/O) ratio (log(N/O)$=-$2.3) is about ten times lower than the bulk of DLAs and H\,II regions of BCGs.

The goal of this article is to perform a uniform and unbiased analysis of nitrogen and (N/O) ratios in a 
sample of Galactic metal-poor stars and compare them with theoretical models and many measurements available 
for extragalactic systems. We employed spectral synthesis of NH band in near-UV high-resolution spectra.
Different spectral features can be used to measure the nitrogen abundance in solar-type stars. Weak high 
excitation ($\chi$=10.34 eV) near-infrared \ion{N}{i} lines at 7468.31, 8216.34, 8683.4, 8703.25 and 8718.83 \AA\ 
disappear at metallicities [Fe/H]$< -$1 and for the analysis of N in metal-poor stars one is left with the CN and NH 
molecular bands centered at 3883 and 3360 \AA, respectively. The first studies of the NH band by Sneden (1973),
Bessell \& Norris (1982) showed that one can use this spectral feature as an independent abundance indicator.  
Observations of the NH band at 3360 \AA\ have allowed many investigators to delineate the Galactic evolution 
of N down to [Fe/H]$\sim -$3 (Laird 1985; Carbon et al. 1987; Tomkin \& Lambert 1984; Israelian, 
Garc\'{\i}a-L\'opez \& Rebolo 2000).
Laird (1985) obtained constant [N/Fe]= $-0.67\pm0.14$ in the metallicity range 
$-2.5 <$ [Fe/H]$< 0$ from the intermediate resolution spectra of 116 stars while Carbon et al.\ (1987) 
found [N/Fe]= $\pm$0.113$\pm$0.063 from a similar-quality spectra of 76 subdwarfs. Unfortunately, NH 
lines are blended with several strong Ti and Sc lines and it is preferable to use high resolution spectra in 
order to avoid any systematics caused by the overabundance of Ti (an $\alpha$-element) and Sc in metal-poor stars. 
Tomkin \& Lambert (1984) used high resolution spectra of 14 metal-poor stars in the range $-2.3<$ [Fe/H]$< -0.3$ 
and found [N/Fe]$\simeq -$0.25.

\section{Observations}

The near-UV spectra of our targets listed in Table 1 were obtained with the UES spectrograph, at the 4.2 m WHT 
(Observatorio del Roque de los Muchachos, La Palma), the UVES spectrograph, at the VLT/UT2 
Kueyen Telescope (ESO,Chile), CASPEC at the 3.6 m ESO telescope (La Silla, Chile) and UCLES at the 3.9 m
AAT (Australia). The spectra for these metal-poor stars have been collected and used to derive Be, O, 
Cu and Zn abundances in a series of papers (e.g. Molaro et al.\ 1987; Israelian et al.\ 1998, 2001; Garc\'{\i}a L\'opez et
al.\ 1995,1998; Bihain et al.\ 2004). The high resolution spectra ($R \ge 50000$  in all the spectra except CASPEC which had
$R \sim 35000$) near NH band 3360 \AA\ have $S/N$ ratios normally higher than 80. The S/N ratio was 
in the range 250--350 in the VLT/UVES spectra of G64-12, G275-4 and LP815-43. The spectra were normalized
using a 5th order polynomial of the CONTINUUM task of IRAF. This normalization was good enough in all
our targets with [Fe/H]$< -1$. However, the spectra of more metal-rich stars were normalized with regard 
to the solar spectrum of Kurucz et al.\ \cite{Kur84} as described in Ecuvillon et al.\ (2004).

\begin{table*}[t]
\caption[]{Nitrogen and oxygen abundances from NH band synthesis and from literature, respectively, for metal-poor stars.}
\begin{center}
%\begin{scriptsize}
\begin{tabular}{lcccccccrr}
\hline
\noalign{\smallskip}
Star & $T_\mathrm{eff}$ & $\log {g}$ & $\xi_t$ & [Fe/H] & [N/H]  & [O/H]  & [N/O] \\
 & (K) & (cm\,s$^{-2}$) & (km\,s$^{-1}$) & & & & & \\
\hline 
\hline
\noalign{\smallskip}
\object{LP\,815-43}          & $6265\pm125$ & $4.54\pm0.30$ & 1.00 & $-2.74$ &   $<-$2.44   &  $-2.11\pm0.27^2$    &$<-$0.33 \\ 
\object{LHS 540}             & $5993\pm71$ & $3.88\pm0.20$ & 1.00 & $-1.48$ & $-1.58\pm0.16$  &  - & - \\
\object{BD\,+23$^\circ$3130} & $5154\pm58 $ & $2.93\pm0.28$ & 1.25 & $-2.66$ & $-2.86\pm0.16$ &   $-1.78\pm0.17^2$    & $-$1.08 \\
\object{BD\,+23$^\circ$3912} & $5734\pm67 $ & $3.83\pm0.13$ & 1.00 & $-1.50$ & $-1.80\pm0.15$ &   $-0.94\pm0.17^3$    & $-$0.86 \\ 
\object{BD\,+26$^\circ$3578} & $6263\pm76 $ & $3.93\pm0.21$ & 1.00 & $-2.36$ &   $<-$1.36     &   $-1.42\pm0.11^3$    &$<-$0.06 \\ 
\object{BD\,+37$^\circ$1458} & $5326\pm54 $ & $3.30\pm0.23$ & 1.25 & $-2.17$ & $-1.97\pm0.15$ &   $-1.36\pm0.21^1$    & $-$0.61\\
\object{G64-12}              & $6318\pm150$ & $4.20\pm0.30$ & 1.00 & $-3.05$ & $-1.90\pm0.25$ &   $-1.88\pm0.31^2$    & $-$0.02 \\ 
\object{G275-4}              & $6212\pm150$ & $4.13\pm0.30$ & 1.00 & $-2.99$ &   $<-$2.29     &   $-2.08\pm0.32^2$    & $<-$0.21 \\ 
\object{HD\,6582}            & $5296\pm67 $ & $4.46\pm0.05$ & 1.00 & $-0.86$ & $-1.05\pm0.14$ &   $-0.53\pm0.16^1$    & $-$0.52 \\
\object{HD\,19445}           & $6095\pm69 $ & $4.45\pm0.05$ & 1.00 & $-2.04$ & $-1.69\pm0.12$ &   $-0.82\pm0.17^1$    & $-$0.87 \\
\object{HD\,64090}           & $5417\pm65 $ & $4.55\pm0.06$ & 1.00 & $-1.74$ & $-1.75\pm0.14$ &   $-1.00\pm0.15^1$    & $-$0.75 \\
\object{HD\,76932}           & $5859\pm72 $ & $4.13\pm0.04$ & 1.00 & $-0.90$ & $-1.12\pm0.15$ &   $-0.56\pm0.18^1$    & $-$0.56 \\
\object{HD\,84937}           & $6277\pm75 $ & $4.03\pm0.09$ & 1.00 & $-2.06$ &   $<-$2.06     &   $-1.29\pm0.17^1$    & $<-$0.77 \\
\object{HD\,87140}           & $5086\pm43$ & $2.96\pm0.31$ & 1.25 & $-1.83$ & $-1.83\pm0.15$    &  -  & - \\
\object{HD\,94028}           & $6058\pm72 $ & $4.27\pm0.06$ & 1.00 & $-1.49$ & $-1.34\pm0.12$   &  $-0.78\pm0.16^1$    & $-$0.56 \\
\object{HD\,103095}          & $5047\pm68 $ & $4.61\pm0.05$ & 1.00 & $-1.43$ & $-1.83\pm0.14$   &  $-1.06\pm0.19^3$    & $-$0.77 \\
\object{HD\,128279}          & $5130\pm66$ & $2.85\pm0.22$ & 1.25 & $-2.11$ & $-2.41\pm0.16$ &  -  & - \\
\object{HD\,134169}          & $5877\pm72 $ & $3.96\pm0.07$ & 1.00 & $-0.81$ & $-0.81\pm0.15$  &  $-0.52\pm0.17^1$    & $-$0.29 \\
\object{HD\,140283}          & $5723\pm67 $ & $3.68\pm0.06$ & 1.00 & $-2.47$ & $-2.42\pm0.11$  &  $-1.58\pm0.15^1$    & $-$0.84 \\
\object{HD\,157214}          & $5715\pm43 $ & $4.21\pm0.03$ & 1.00 & $-0.35$ & $-0.45\pm0.12$  &  $-0.24\pm0.11^1$    & $-$0.21 \\
\object{HD\,165908}          & $5905\pm42 $ & $4.11\pm0.03$ & 1.00 & $-0.67$ & $-0.77\pm0.12$  &  $-0.63\pm0.10^1$    & $-$0.14 \\
\object{HD\,166913}          & $6181\pm74 $ & $4.18\pm0.07$ & 1.00 & $-1.68$ & $-1.38\pm0.12$  &  $-0.69\pm0.18^1$    & $-$0.69 \\
\object{HD\,170153}          & $5954\pm42 $ & $4.13\pm0.03$ & 1.00 & $-0.65$ & $-0.86\pm0.12$  &  $-0.74\pm0.12^1$    & $-$0.12 \\
\object{HD\,188510}          & $5559\pm65 $ & $4.54\pm0.06$ & 1.00 & $-1.57$ & $-1.57\pm0.14$  &  $-1.01\pm0.17^1$    & $-$0.56 \\
\object{HD\,189558}          & $5681\pm70 $ & $3.81\pm0.08$ & 1.00 & $-1.12$ & $-1.32\pm0.15$  &  $-0.74\pm0.17^1$    & $-$0.58 \\
\object{HD\,194598}          & $6029\pm77 $ & $4.28\pm0.07$ & 1.00 & $-1.20$ & $-1.10\pm0.15$  &  $-0.68\pm0.19^1$    & $-$0.42 \\
\object{HD\,201889}          & $5618\pm67 $ & $4.04\pm0.08$ & 1.00 & $-0.95$ & $-1.05\pm0.14$  &  $-0.80\pm0.20^1$    & $-$0.25 \\
\object{HD\,201891}          & $5916\pm77 $ & $4.26\pm0.05$ & 1.00 & $-1.05$ & $-0.95\pm0.15$  &  $-0.60\pm0.18^1$    & $-$0.35 \\
\object{HD\,218502}          & $6182\pm77$ & $4.08\pm0.08$ & 1.00 & $-1.76$ & $<-$2.06 &   - & - \\
\object{HD\,211998}          & $5282\pm180$ & $3.28\pm0.28$ & 1.25 & $-1.48$ & $-1.98\pm0.30$  &  $-1.27\pm0.38^1$    & $-$0.71 \\
\object{HD\,225239}          & $5528\pm44 $ & $3.74\pm0.15$ & 1.00 & $-0.50$ & $-0.65\pm0.09$  &  $-0.71\pm0.12^1$    & 0.06 \\
\noalign{\smallskip}
\hline
\end{tabular}
%\end{scriptsize}
\end{center}
\footnotetext{}{$^1$ from Israelian et al.\ (1998)}\\
\footnotetext{}{$^2$ from Israelian et al.\ (2001)}\\
\footnotetext{}{$^3$ from Boesgaard et al.\ (1999)}\\
\label{tab1}
\end{table*}

\begin{table*}[t]
\caption[]{Nitrogen and oxygen abundances from NH band synthesis (Ecuvillon et al.\ 2004) and from literature (last
column), respectively, for metal-rich stars.}
\begin{center}
%\begin{scriptsize}
\begin{tabular}{lcccccccrr}
\hline
\noalign{\smallskip}
Star & $T_\mathrm{eff}$ & $\log {g}$ & $\xi_t$ & [Fe/H] & [N/H]  & [O/H]  & Ref. \\
 & (K) & (cm\,s$^{-2}$) & (km\,s$^{-1}$) & & & & & \\
\hline 
\hline
\noalign{\smallskip}
\object{HD\,52265} & 6105 & 4.28 & 1.36 &  0.23  & 0.25     &  0.02    & 1 \\
\object{HD\,75289} & 6143 & 4.42 & 1.53 &  0.28  & 0.21     & -0.05    & 1 \\
\object{HD\,82943} & 6015 & 4.46 & 1.13 &  0.30  & 0.39     &  0.21    & 1 \\
\object{HD\,83443} & 5454 & 4.33 & 1.08 &  0.35  & 0.26     &  0.44    & 1 \\
\object{HD\,9826} & 6212 & 4.26 & 1.69 &  0.13  & 0.10      & -0.07    & 2 \\
\object{HD\,22049} & 5073 & 4.43 & 1.05 & -0.13  & -0.05    &  0.00    & 2 \\
\object{HD\,38529} & 5674 & 3.94 & 1.38 &  0.40  & 0.42     & -0.01    & 2 \\
\object{HD\,92788} & 5821 & 4.45 & 1.16 &  0.32  & 0.42     &  0.08    & 3 \\
\object{HD\,120136} & 6339 & 4.19 & 1.70 &  0.23  & 0.20    &  0.08    & 2 \\
\object{HD\,134987} & 5776 & 4.36 & 1.09 &  0.30  & 0.40    &  0.22    & 4 \\
\object{HD\,143761} & 5853 & 4.41 & 1.35 & -0.21  &-0.30    & -0.12    & 2 \\
\object{HD\,209458} & 6117 & 4.48 & 1.40 &  0.02  & 0.01    & -0.07    & 3 \\
\object{HD\,217014} & 5804 & 4.42 & 1.20 &  0.20  & 0.20    &  0.16    & 4 \\
\object{HD\,217107} & 5645 & 4.31 & 1.06 &  0.37  & 0.40    &  0.17    & 2 \\
\object{HD\,222582} & 5843 & 4.45 & 1.03 &  0.05  & 0.12    & -0.04    & 4 \\
\noalign{\smallskip}
\hline
\end{tabular}
%\end{scriptsize}
\end{center}
\footnotetext{}{$^1$ from Santos et al.\ (2000)}\\
\footnotetext{}{$^2$ from Takeda et al.\ (2001)}\\
\footnotetext{}{$^3$ from Sadakane et al.\ (2002)}\\
\footnotetext{}{$^3$ from Gonzalez et al.\ (2001)}\\
\label{tab2}
\end{table*}

\section{Analysis}
\subsection{Rationale}

Our goal is to perform a uniform and unbiased analysis of the (N/O) ratio. Abundances of these elements 
can be determined from several molecular and atomic features. Both elements have high excitation atomic lines
in the near IR and low excitation diatomic molecular bands in the near-UV. Many authors have compared abundances
from \ion{O}{i} against OH and \ion{C}{i} against CH (Israelian et al.\ 1998, 2001; Boesgaard et al.\ 1999;
Nissen et al.\ 2002; Tomkin et al.\ 1992) in metal-poor stars. While some authors found good agreement 
and others did not, it was realized that such comparisons are very sensitive to the stellar parameters. 
In fact, both O and C have abundance indicators that are less sensitive to the stellar parameters and almost 
independent on the non-LTE effects (i.e. [\ion{O}{i}] 6300 and [\ion{C}{i}] 8727 \AA). A careful comparison 
of all three abundance indicators of O has been carried out for only few stars and positive results 
have been obtained (Israelian et al.\ 2001; Nissen et al.\ 2002). No detailed analysis has ever been carried
out for Nitrogen because the high excitation lines of \ion{N}{i} disappear at
metallicities below [Fe/H]$< -1$. We note a good agreement found between the near-UV NH band and the high 
excitation \ion{N}{i} 7468 \AA\ line in metal-rich stars (Ecuvillon et al.\ 2004). Even armed with
different abundance indicators and high quality observations, we often face situations where little
can be done to resolve apparent abundance conflicts (e.g. Takeda 2003; Fulbright \& Johnson 2003; 
Israelian et al.\ 2004). New generation 3D models of atmospheres were called in to save the situation 
(Asplund \& Garcia-Perez 2001; Asplund et al.\ 2003). These models suggest that diatomic molecules 
(e.g. OH, CH and NH) in the near UV strongly overestimate elemental abundances derived using 1D models. The effect is 
strong in hot and very metal-poor subdwarfs.
However, 3D models do not provide consistent abundances from 
the near-UV OH, \ion{O}{i} 7771--5 \AA\ and [\ion{O}{i}] 6300 \AA\ lines (Nissen et al.\ 2002). These models are not 
calibrated and no grid is available to check them against observations of many solar-type stars with 
different atmospheric parameters. 
The models have yet to demonstrate that they can explain various observational characteristics (such as 
Balmer line profiles, continuum energy distribution, consistent abundances from different abundance indicators) in many 
stars with different parameters. 

To ensure a homogeneous study of the (N/O) ratio from the NH and OH features in the near-UV, 
we used the same model atmospheres and tools as in our previous studies (e.g. Israelian et al.\ 1998, 2001).
 The oxygen abundances were compiled from the papers of Israelian et al. (1998, 2001) and Boesgaard et al. (1999).
Few stars have been added to the lists of Israelian et al.\ (1998, 2001) while the stellar parameters 
and [Fe/H] (Table~1) were taken from Bihain et al.\ (2004). The effective temperatures were computed
by Bihain et al.\ (2004) 
using the IRFM calibration versus $V$--$K$ colors, the gravities from {\it Hipparcos} parallaxes and the 
metallicities from the literature. The parameters of G64-12, G275-4 and LP815-43 were taken 
from Israelian et al.\ (2001). Uncertainties in the atmospheric parameters are of the order of 80 K in $T_\mathrm{eff}$, 
0.1 in $\log{g}$ and 0.15 in the metallicity. The effect of $T_\mathrm{eff}$, $\log\mathrm{g}$ and [Fe/H] on the 
NH abundance was considered by Ecuvillon et al.\ (2004). We found that the effect is not different in 
metal-poor stars and assumed the same sensitivity values as in Ecuvillon et al.\ (2004). The sensitivity of NH 
abundance was applied to the propagation of the error of each parameter error on abundances. A continuum uncertainty effect 
of the order of 0.1\,dex was considered in all targets except for three stars observed with VLT/UVES where the error 
was less than 0.05\,dex. All errors were added in quadrature to obtain final uncertainties for nitrogen 
abundances listed in Table~\ref{tab1}.

\begin{figure*}
\centering
\includegraphics[height=25cm]{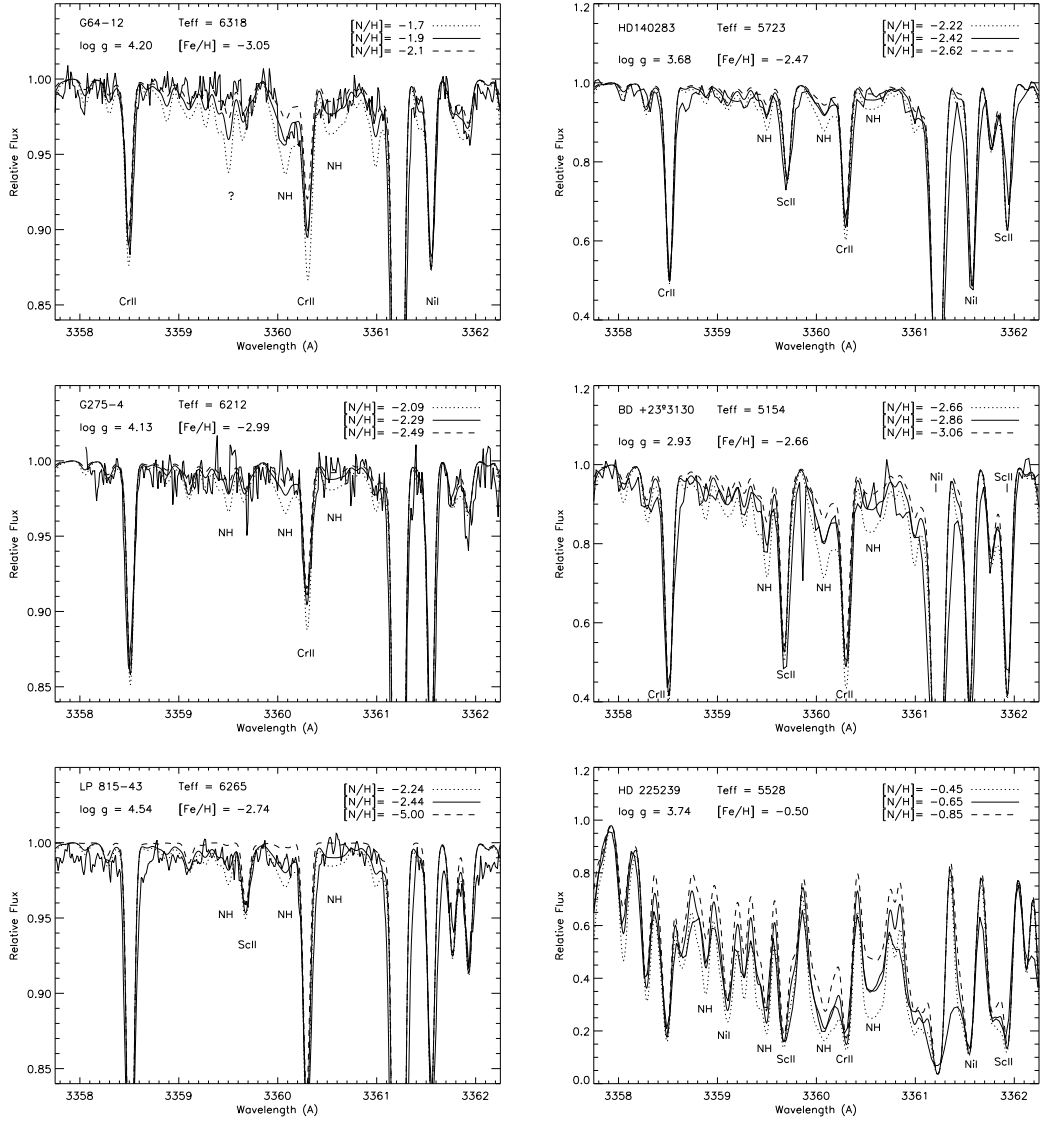}
\caption{Observations and synthesis of the NH band in the high-resolution near-UV spectra of several 
metal-poor stars.}
\label{fig1}
\end{figure*}

%\begin{figure*}
%\begin{center}
%\includegraphics[height=6cm]{grafG6412.eps}
%\includegraphics[height=6cm]{grafHD140283.eps}
%\includegraphics[height=6cm]{grafG2754.eps}
%\includegraphics[height=6cm]{grafG17047.eps}
%\includegraphics[height=6cm]{grafLP81543.eps}
%\includegraphics[height=6cm]{grafHD225239.eps}
%\end{center}
%\caption[Observations and synthesis of the NH band in the high-resolution near-UV spectra of several 
%metal-poor stars.]{Observations and synthesis of the NH band in the high-resolution near-UV spectra of several 
%metal-poor stars.}
%\label{fig1}
%\end{figure*}

\subsection{Synthesis of NH band}

Nitrogen abundances were determined by fitting an LTE synthetic spectrum to data in the region $\lambda\lambda$ 3345-3375 
\AA. A detailed line list from Ecuvillon et al.\ (2004) was employed. These authors have slightly modified the line list 
of Yakovina \& Pavlenko (1998) without applying any change to the continuum level when fitting the 
high resolution Kurucz Solar Atlas (Kurucz et al. 1984) with a solar model having $T_\mathrm{eff}= 5777\,K$, 
$\log\mathrm{g}=4.44$ and $\xi_t=1.0\,km\,s^{-1}$. To measure the N abundance in stars with $-1<$[Fe/H]$<$0 we used 
the program FITTING\footnote{FITTING is developed by Jonay I. Gonz\'alez-Hern\'andez (Instituto de Astrof\'{\i}sica de Canarias).}.
This program creates a grid of synthetic spectra computed with MOOG (Sneden 1974) for several free parameters (e.g. 
abundances, atmospheric parameters, etc...) and compares them with the observed spectrum. The best fit is determined by 
applying a $\chi^2$ minimization method to the spectral regions considered most significant (Ecuvillon et al.\ 2004). 
Instrumental broadening was computed using a Gaussian function corresponding to the resolution of the instrument, 
while the value of $v\sin{i}$ was adjusted by eye and was always less than 3 km\,s$^{-1}$ since all our targets are evolved
slow rotators. Our tests with the $\chi^2$ technique show that the nitrogen abundance (as well as the abundances 
of all other elements) is not affected by $v\sin{i}$. The fits were performed by scaling the abundances of all elements to 
the [Fe/H] value and then modifying them typically within $\pm$0.4\,dex until the best solution was found. It was much 
easier to derive the nitrogen abundance in stars with [Fe/H]$< -$2, where most of the
blends disappear and the position of continuum can be determined with better precision. The goodness of the spectral 
synthesis in these stars was estimated by eye. 
 In figure~\ref{fig1} we show the results of the fits for six of our program stars which span the full 
range of metallicities covered by our sample. In each panel the observed spectrum of a star is compared with 
the best fitting synthetic spectrum and with two other synthetic spectra with different N abundance 
which bracket the best fitting spectrum. Such plots allow to estimate the errors associated to the N abundance
measurement. It is noticeable in this plot  that the most metal--poor star in our sample, \object{G64-12}, has  
a clearly measurable NH band. As will be stated in a well defined sense in Section 4.4, this star appears to 
be N-rich ([N/Fe]=+1.15 ). Stars \object{LP815-43} and \object{G275-4} (first and second left panel from the bottom)
form an interesting pair: their atmospheric parameters are very similar, yet the NH band is measurable in the 
former, but not in the latter. Neither of these stars is known to be in a binary system.
\object{LP815-43} ([N/Fe] = + 0.63), like \object{G64-12}, belongs to the group of N-rich stars.
From the upper right panel one can see that we can confirm the detection of NH band in \object{HD\,140283} 
(Bessell \& Norris 1982; Tomkin \& Lambert 1984). From the spectrum of \object{HD\,225239} (lower right panel)
it appears clear how the analysis of the NH is more difficult at higher metallcities due to the
strong blending with metallic lines and saturation of the NH lines.

\begin{figure*}
\centering
\includegraphics[height=6cm]{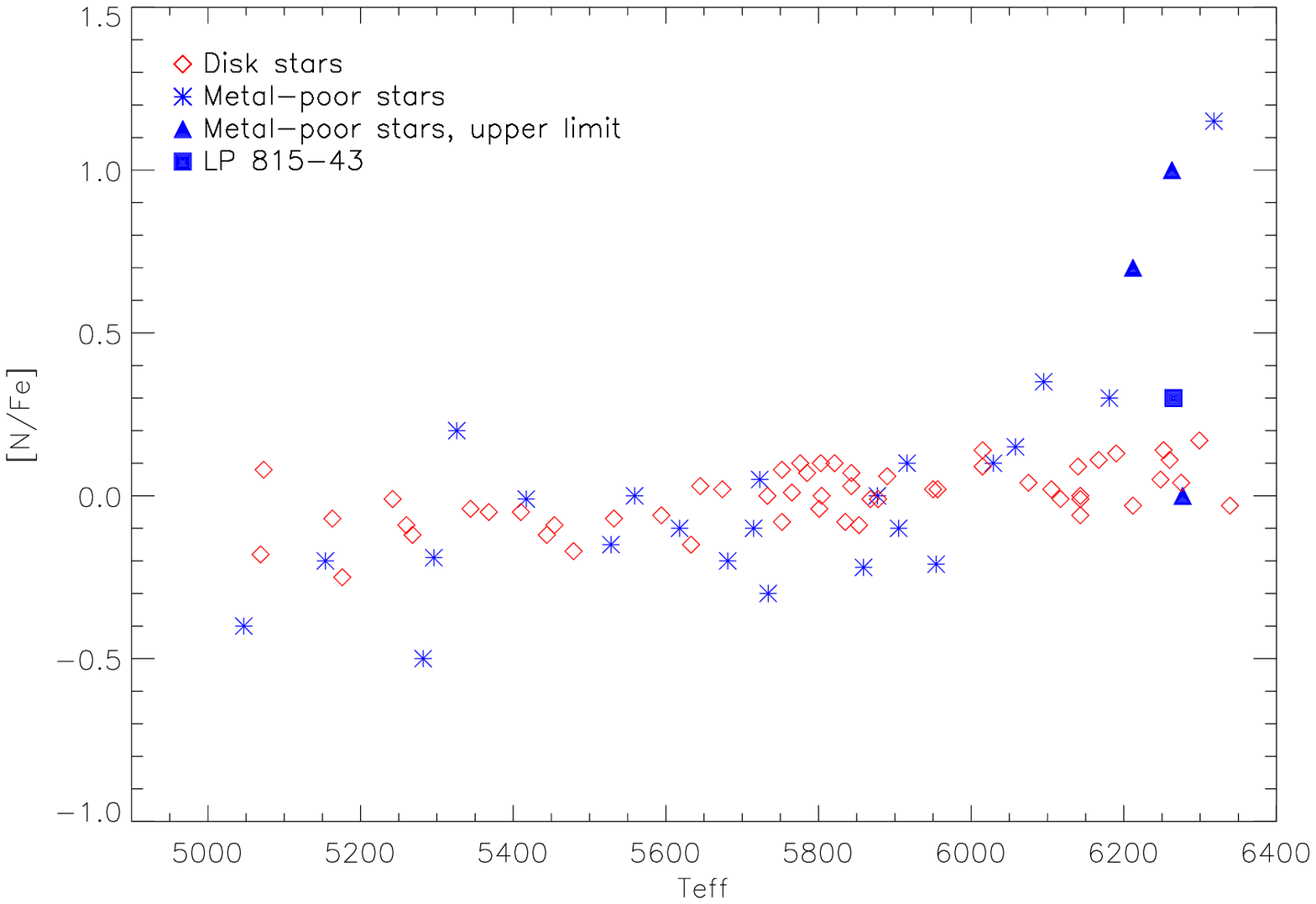}
\includegraphics[height=6cm]{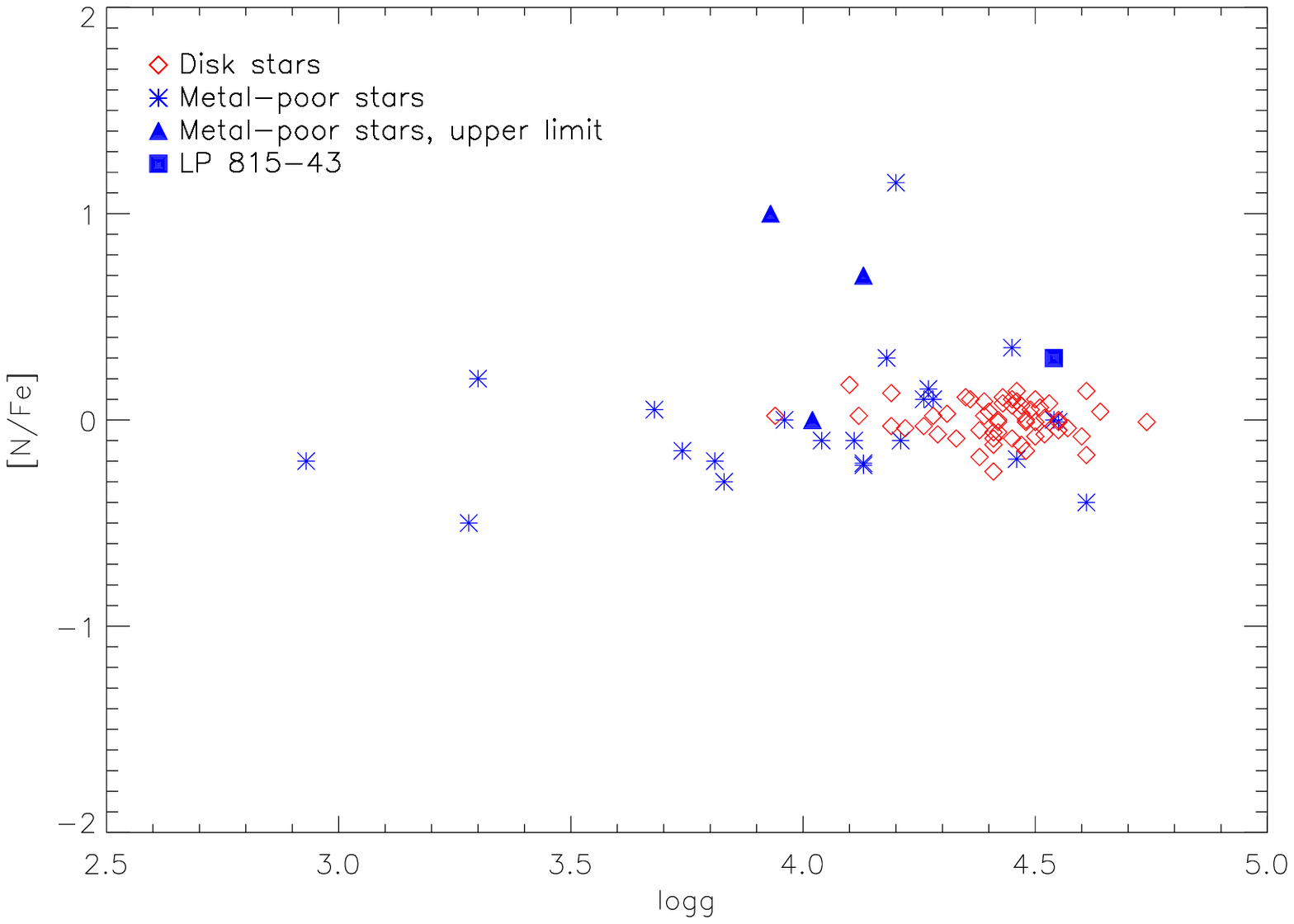}
\includegraphics[height=6cm]{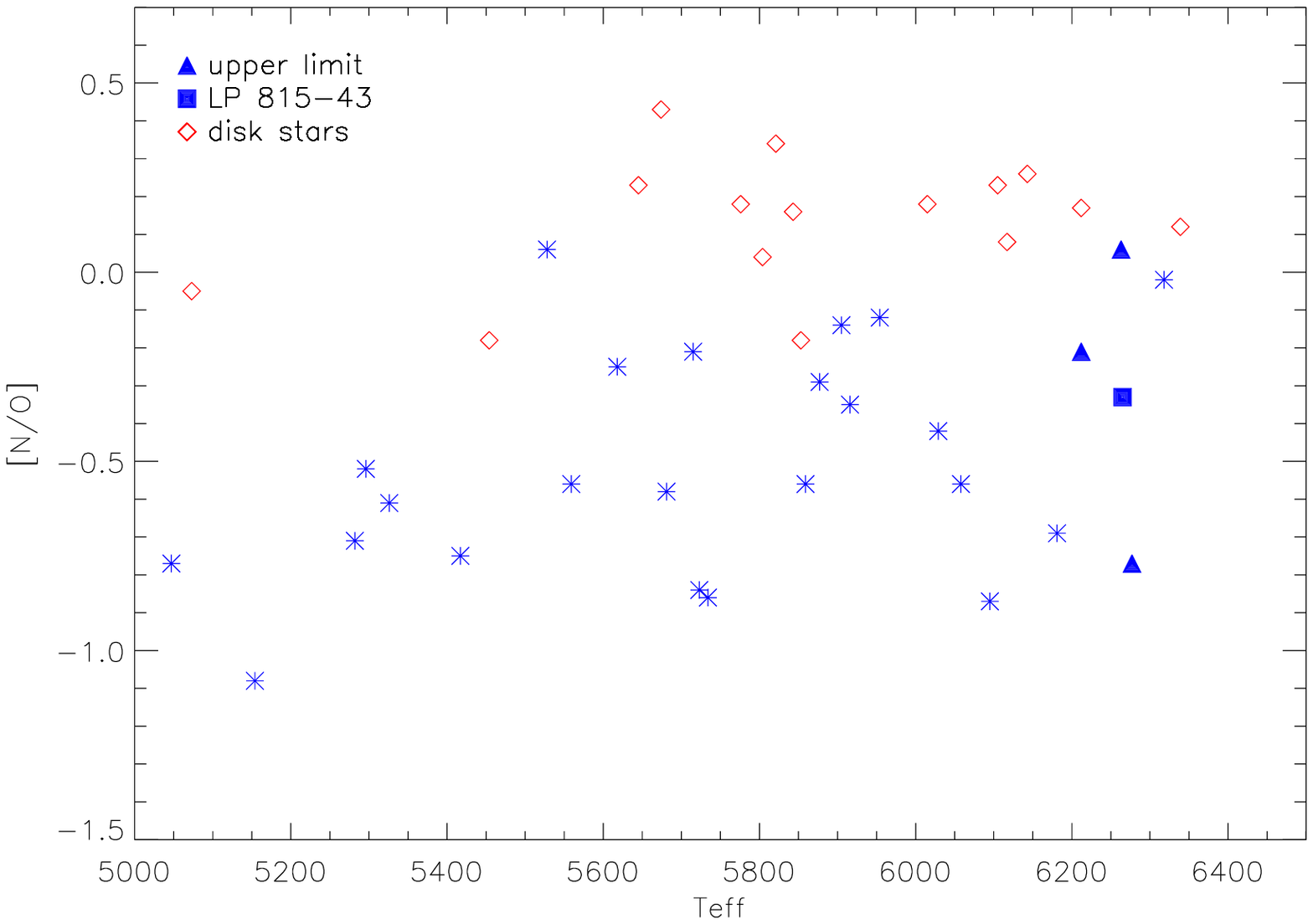}
\includegraphics[height=6cm]{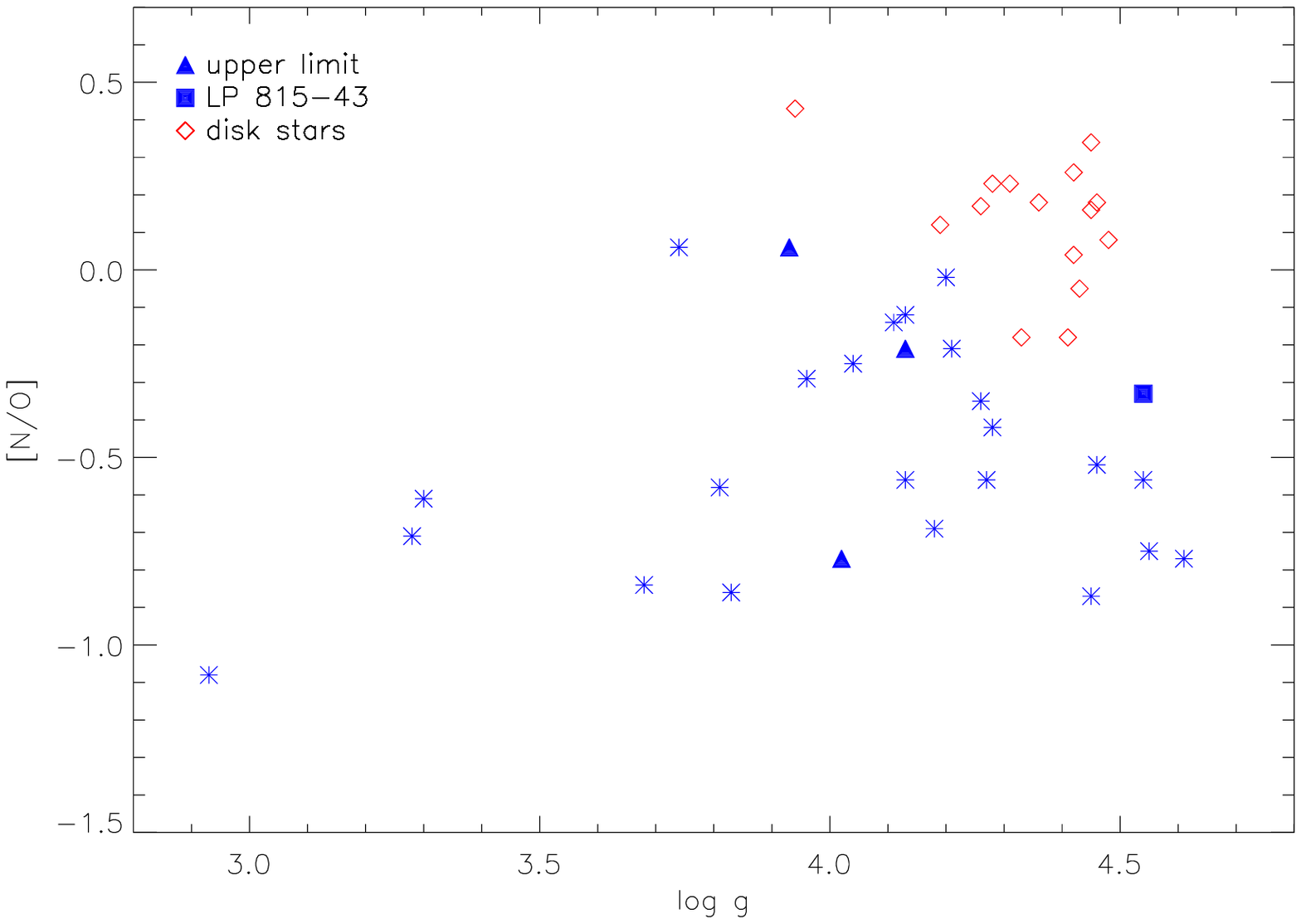}
\caption{[N/Fe] and [N/O] ratios versus $T_{\rm eff}$ and $\log{g}$.}
\label{fig2}
\end{figure*}

The solar abundance of nitrogen was assumed log $\epsilon$(N)=8.05 (on the customary scale in which log $\epsilon$(H)=12) 
from Anders \& Grevesse (1989). There is, however, some debate over the oxygen abundance in the Sun 
(Allende Prieto et al.\ 2001; Nissen et al.\ 2002; Asplund et al.\ 2003). The latest value, log $\epsilon$(O)=8.66, from 
3D models is about 0.3\,dex smaller than the original value from Anders \& Grevesse (1989). The solar oxygen abundance 
used for the differential analysis of OH lines by Israelian et al.\ (1998) was log $\epsilon$(O)=8.93 from 
Anders \& Grevesse (1989). In order to be consistent with the solar $gf$ values used by Israelian et al.\ (1998) and with 
1D model analysis in general, we have decided to use the same value in this article. This  assumption will
not introduce any errors since one can always switch to the scale where log $\epsilon$(O)=8.66 or any other value 
discussed in the literature and accordingly modify solar $gf$ values of the OH lines from Israelian et al.\ (1998). 
The oxygen abundances in our targets come from the near-UV OH lines analysed by Israelian 
et al.\ (1998, 2001) and Boesgaard et al.\ (1999). The oxygen abundances listed in these articles were updated for the 
parameters listed in the Table~\ref{tab1}, using the sensitivity of OH to  $T_\mathrm{eff}$, $\log\mathrm{g}$ and 
[Fe/H] given in Israelian et al.\ (1998). 

We analysed near-UV high-resolution spectra of 31 metal-poor stars with the goal of delineating Galactic trends 
of [N/Fe], [N/H] and [N/O]. All our targets are unevolved metal-poor stars with [Fe/H] values between $-0.3$ and 
$-3.1$ and $T_\mathrm{eff}$ between 5000 and $6300\,\mathrm{K}$. The results of our analysis are listed in Table~\ref{tab1}. 
We have investigated the presence of the possible systematic dependence of [N/Fe] and [N/O] on $T_{\rm eff}$ and $\log g$ 
in Fig.~\ref{fig2}. The absence of any trend in these figures suggests that the uncertainties in the derived oxygen 
abundances are only related to the errors listed in Table~\ref{tab1}. On the whole, the [N/Fe] vs. [Fe/H] trend appears to 
be almost flat with a slope of $0.01\pm0.02$ (without considering G64-12, LP815-43 and the upper limits).
In Fig. 3 we plot the [N/O] ratios in our progran stars as a function of [O/H], [Fe/H] and [N/H]. 
Like in Fig. 2 the diamond shapes are the N measurements from Ecuvillon et al. (2004). From all the three
plots it is clear that the two stars G 64-12 and LP 815-43 have N/O ratios {\em higher} than other stars of 
similar metallicity, which justify their designation as N-rich. What is also apparent is that there is an 
increase in the N/O ratio with metallicity. To highlight this increase we plotted, as a dashed line, in each panel a linear
regression, the slope of which  is given at the bottom of each plot. In the two upper panels we also plotted, as a reference,
a horizontal dotted line corresponds to [N/O]=-0.62, which is the level of the plateau in N/O observed in 
Blue Compact Galaxies (Izotov \& Thuan, Henry et al. 2000). Figure  4 shows that the [N/Fe] ratio is constant
whichever the metallicity,  and that [N/H] increases linearly with [Fe/H]. The inference from these plots is
that  nitrogen and iron vary in lockstep, in substantial agreement with the main conclusion of Carbon et al. (1987).

In order to extend our plots to the regime of metal-rich stars we used NH abundance from Ecuvillon et al.\ (2004)
and oxygen abundances from the literature (Santos, Israelian \& Mayor 2000; Takeda et al.\ 2001; Gonzalez et al.\ 2001; 
Sadakane et al.\ 2002). Oxygen abundances in these studies were obtained either from the triplet 
at 7771-5 \AA\ or from the forbidden line at 6300 \AA. We do not think this will introduce large errors in the final (N/O) 
ratios since different abundance indicators provide quite consistent abundances within $\pm$0.2\,dex 
in solar metallicity dwarfs. This is true for nitrogen (Ecuvillon et al.\ 2004) and oxygen (Takeda 2003; 
Nissen et al.\ 2002). The compiled data are provided in Table~\ref{tab2}. 

We stress that possible abundance differences from \ion{N}{i} and NH lines produced by uncertainties in the near-UV 
continuum, 3D and other effects are not critical for the present analysis. Such differences cancel out when forming 
abundance ratios from lines formed in the same atmospheric layer. Similar arguments led Tomkin \& Lambert (1984) to derive 
(N/C) ratio from the $\lambda$3360 NH and $\lambda$4300 CH bands. Since the CH, NH and OH hydrids have similar 
dissocitation energies and line formation mechanisms, the derived CNO abundance ratios are insensitive to atmospheric 
parameters and structure. Thus, we measure the (N/O) ratio using the lines of NH and OH in the near UV.

\begin{figure*}
\centering
\includegraphics[height=6cm]{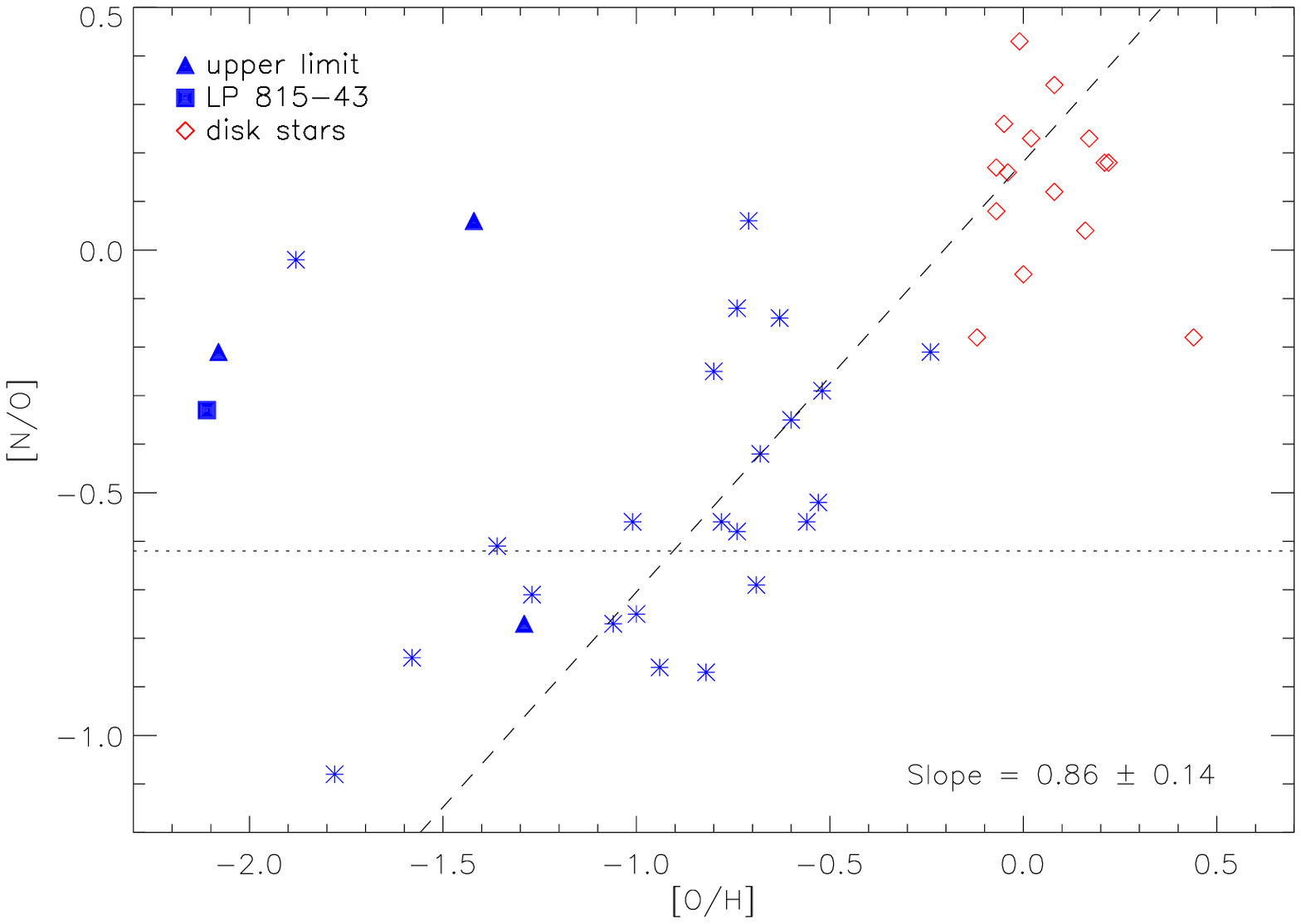}
\includegraphics[height=6cm]{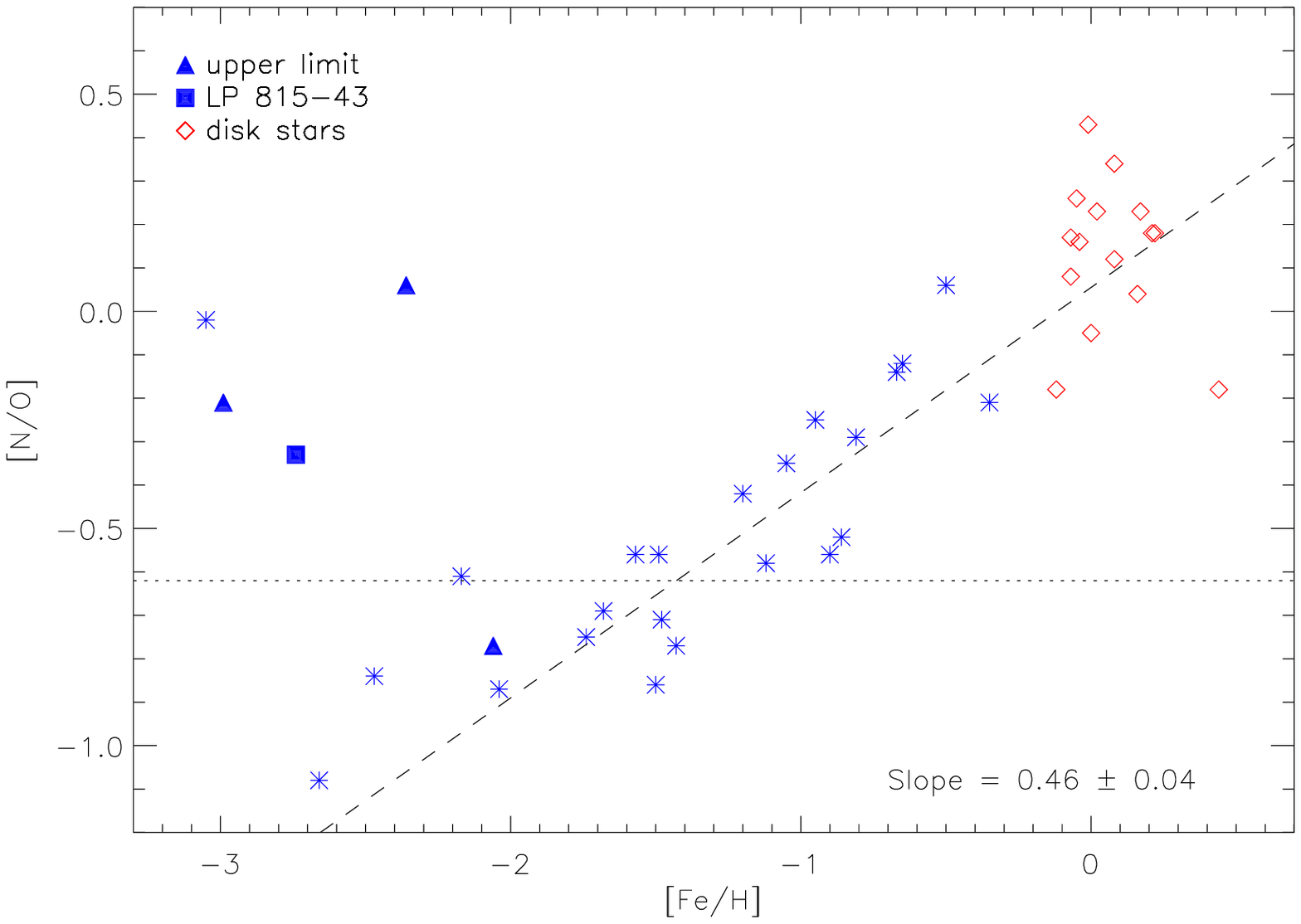}
\includegraphics[height=6cm]{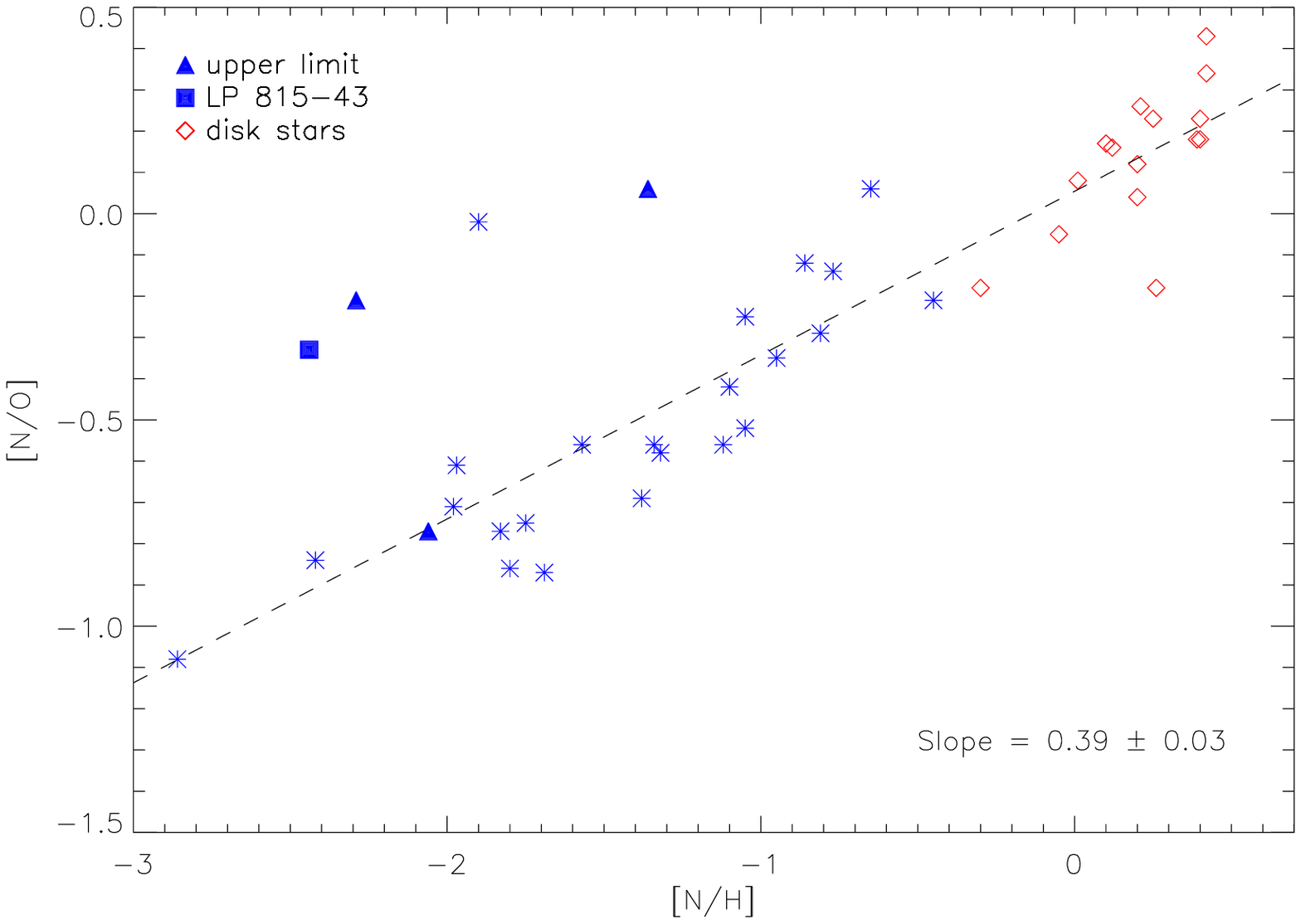}
\caption{The [N/O] ratio as a function of [O/H], [Fe/H] and [N/H] for the programme stars
and for the metal-rich disc stars of Ecuvillon et al.\ (2004).  The dotted line represents 
log(N/O)$\simeq -$1.5 from Henry et al. (2000) while the dashed line is a least-square fit to the
data without taking into account N-rich stars (see the text). A preliminary detection of N is reported in 
LP815-43.}
\label{fig3}
\end{figure*}

\begin{figure*}
\centering
\includegraphics[height=6cm]{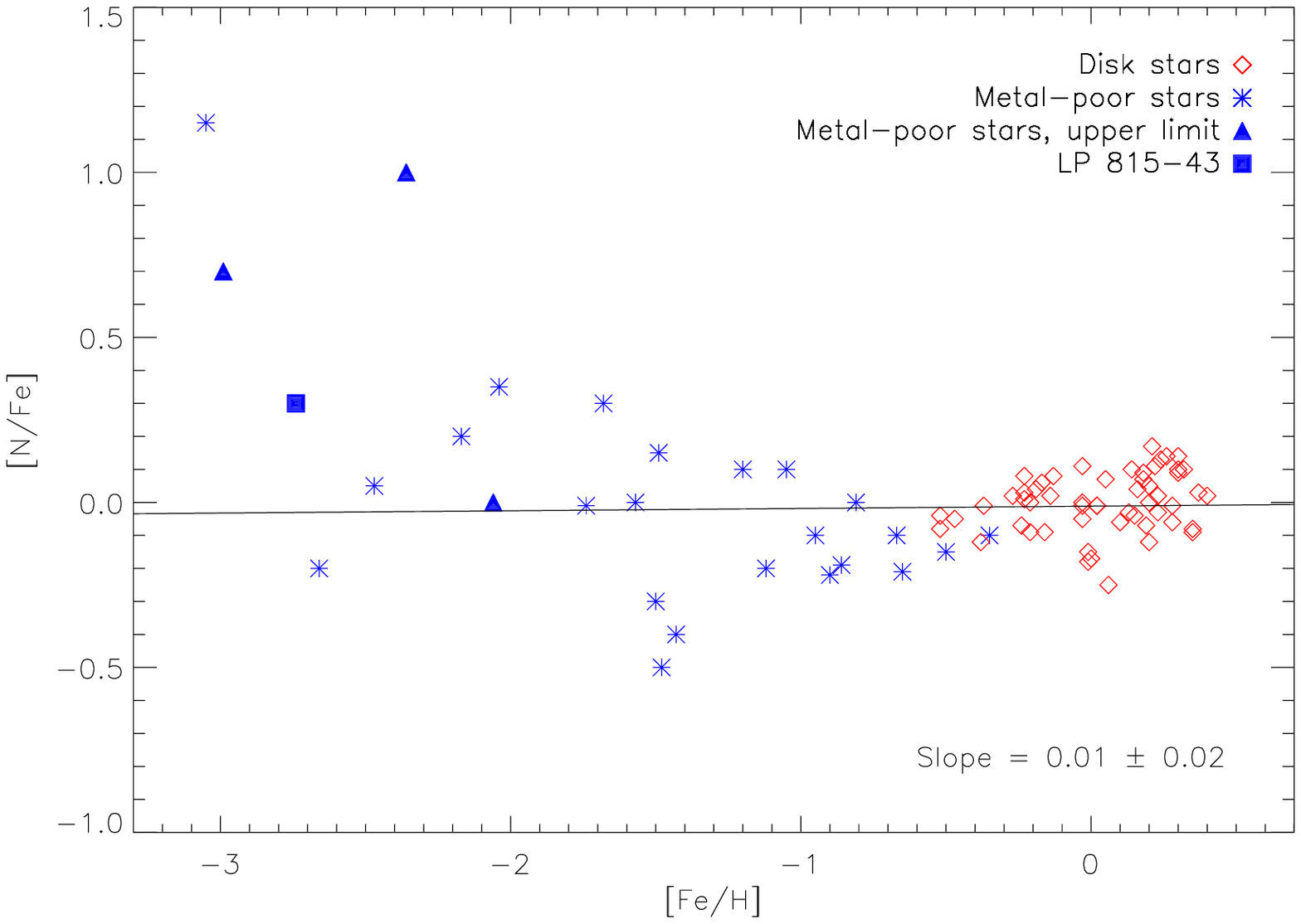}
\includegraphics[height=6cm]{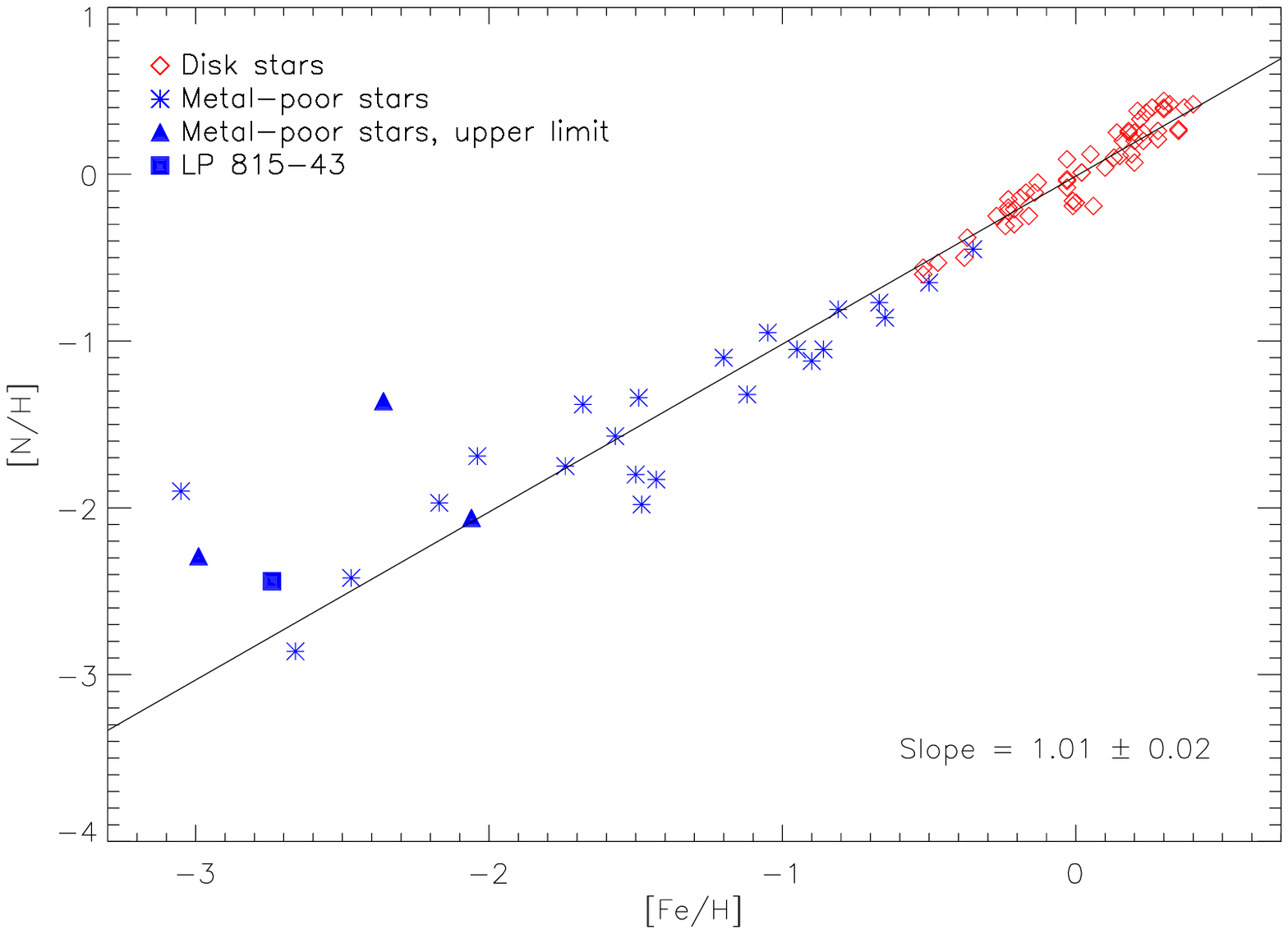}
\caption{[N/Fe] and [N/H] ratios against [Fe/H].}
\label{fig4}
\end{figure*}

\section{Discussion and Conclusions}

\begin{figure*}
\centering
%\resizebox{\hsize}{!}{\includegraphics[clip=true]{pl_models.eps}}
\includegraphics[height=10cm]{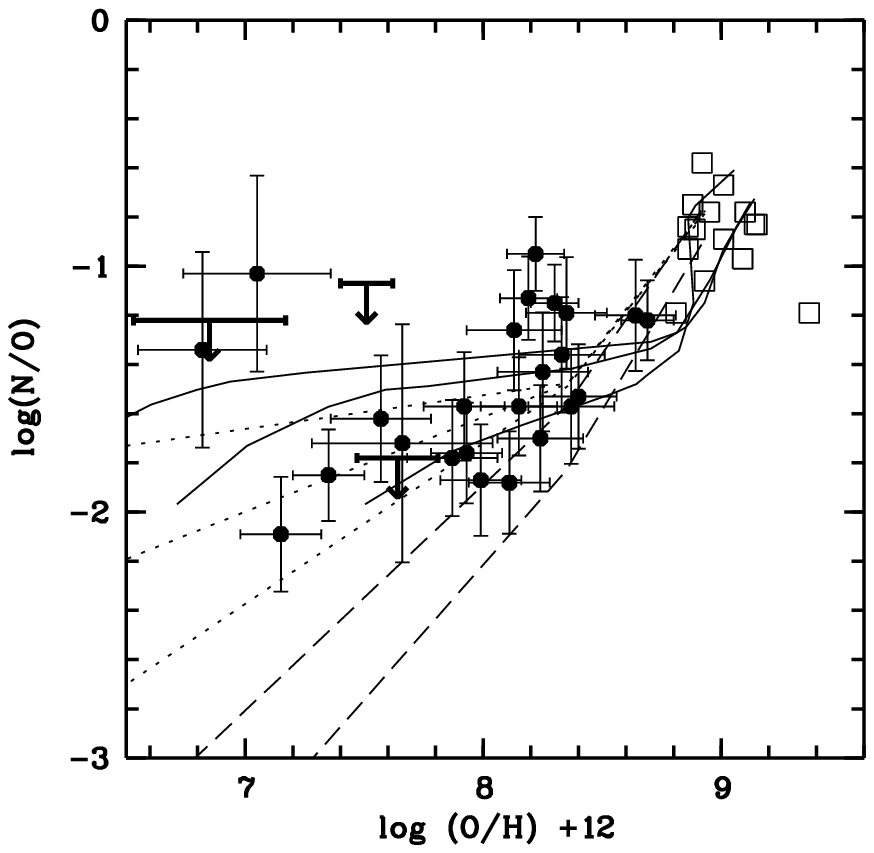}
\caption{Comparison of N/O ratios for programme stars (filled circles) with the simple models of 
Meynet \& Maeder (2002), both including stellar rotation (dotted lines) and with no 
stellar rotation (dashed lines) and with the analytical models of 
Henry et al.\ (2000, solid lines) which were computed to fit the N/O ratios observed in
extragalactic H\,II regions (blue compact galaxies). 
Empty boxes are metal-rich stars listed in Table 2 discussed in the text.}
\label{fig5}
\end{figure*}

\subsection{Comparison with Simple Models}

From our data we can make  a few statements that are quite robust
in the sense that errors in the data cannot invalidate them:
\begin{enumerate}
\item The N/O ratio has an increasing trend together with both oxygen and nitrogen abundance (Fig.~\ref{fig3}).

\item The nitrogen abundance varies in lockstep with that of iron (Fig.~\ref{fig4}).
       
\item There are a few stars that have a N/O ratio higher
      than expected from this trend and their oxygen or
      nitrogen abundance. We shall refer to these stars as N-rich stars.

\end{enumerate} 

Let us now try to see if from these facts we may reach some conclusions with regard to the
nucleosynthesis of nitrogen. Keeping in mind what has been said above concerning N nucleosynthesis, let us compare
our data with some simple models. In Fig.~\ref{fig5} we plot our data and the simple closed box models
of Meynet \& Maeder (2002) and the analytical models of Henry et al.\ (2000). The lower envelope 
of the data seems to lie in the region that could be adequately described by the models without 
any effect of stellar rotation. The remaining data, however, seem to require models
with rotation to be explained. At low metallicities the Galactic stars clearly deviate
from the (N/O) plateau which characterizes extragalactic H\,II regions (i.e. BCGs)
and is described by the middle curve of the three analytical models of Henry et al.\ (2000). 
The data however could be described by the Henry et al.\ (2000) with a rather high star 
formation efficiency (the lowest of the three curves). Since both the Henry et al.\ (2000) 
and  Meynet \& Maeder (2002) models include nitrogen of both primary and secondary origin,
 we believe it is safe to conclude that some primary N production is necessary to explain 
the data. By looking at the models of Meynet \& Maeder (2002) alone it might be tempting to 
conclude that inclusion of stellar rotation is necessary. However this is not necessarily so; 
in fact in the models of Henry et al.\ (2000) the yields for massive stars are taken from 
Maeder (1992), which do not include rotation. It is beyond the scope of this paper to make 
a detailed comparison with chemical evolution models; however, we believe that it is quite likely 
that any model with both primary and secondary nitrogen production would be able of reproducing 
the low metallicity data simply by tuning star formation and infall.
In Fig.~\ref{fig5} there is a group of stars around log(O/H)+12 = 8.2, log(N/O)=$-$1.1  which seems
to stand {\em above} all the models, and, although the errors are such that it is not 
inconsistent with the models.

The comparison with the models also suggests that there might be some intrinsic dispersion in N/O ratios
at any given O abundance. The existence of such a dispersion would support the fact that AGB stars are the 
main contributor of primary nitrogen at low metallicities (Pagel \& Edmunds 1981; Pettini et al.\ 2002).
This would also would support the role of rotation in N production; in fact, different N yields are predicted 
for different starting rotational velocities. Therefore, scatter in N/O ratios would naturally arise from a 
distribution of initial rotational velocities. Differences in star formation history will introduce a scatter 
as well. The errors on the data, however, are too large to 
decide whether such an intrinsic dispersion exists. A mild hint in this direction
comes from the inspection  of the behaviour of the N/O ratio as a function of O and N (Fig.~\ref{fig3}), 
the scatter about the fitting line is larger in the [N/O] vs. [O/H] plane
than in the [N/O] vs. [N/H] plane. If the scatter were caused purely by observational error, it should 
be the same for both planes. If, on the other hand, there were intrinsic scatter caused by rotation, 
the [O/H] would not depend on rotation while [N/O] would, and this plane shows the full effect of this scatter.
Instead {\em both} [N/H] {\em and} [N/O] depend on rotation in a similar manner and the resulting
scatter is smaller in this plane.

\subsection{Comparison with DLAs}

\begin{figure*}
\centering
\includegraphics[height=10cm]{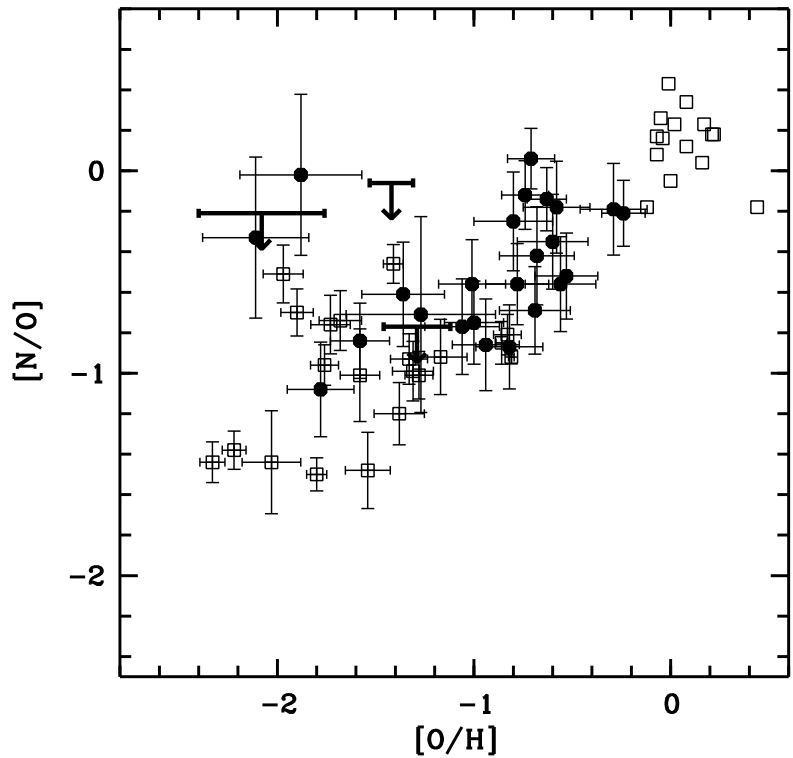}
\caption{The [N/O] as a function of [O/H] for the programme stars (filled
circles) and for the damped Lyman$\alpha$ systems (DLAs)
from Centurion et al. (2003, crossed squares).
We chose to show [N/Si] rather than [N/O] since there are more data points available and the measurements of Si are more
reliable than those of O. However, we should bear in mind that different groups of objects (e.g. DLAs, BCGs and Galactic
stars) may have different star formation histories.}
\label{fig6}
\end{figure*}

A large data set of N abundances in damped Lyman$\alpha$
systems (DLAs) has recently been presented by Centuri\'on et al.\ (2003). In Fig.~\ref{fig6}
we compare our data with that of Centuri\'on et al.\ (2003, refer to that paper for the original 
references to all the data). What is plotted is [N/Si] since the Si measurements in DLAs are more reliable than O
measurements. The [N/O] ratios of DLAs seem to lie in the same
region occupied by Galactic stars with two differences: i) the DLAs extend to lower [N/O]
ratios than galactic stars; ii) although there are some DLAs with ``high'' N/O ratios they never reach
the very high values displayed by a few Galactic stars at comparable metallicity. 
The conclusion is that the N production history is similar in our Galaxy and DLAs, 
but different in BCGs.

Among the DLAs there is a group of 5 which have a very low [N/Si] ($\sim -1.4$) 
and Molaro et al.\ (2003) claim that these represent a second {\em plateau } in nitrogen abundance.
The proposed interpretation of this is that these galaxies are very young and are observed
{\em before } intermediate mass stars had time to contribute to the N abundance.
If our Galaxy has passed such a phase there should be some stars with such a low [N/$\alpha$];
moreover, no star with a lower [N/$\alpha$] ratio should be observed. The lowest point with [N/O]=$-1.08$ corresponds to 
the star BD\,+23$^{\circ}$3130 and is therefore above the ``low N/$\alpha$'' group of 
DLAs, although the errors in abundances (of both the stars and the DLAs) are such that the two are consistent at less
than 1$\sigma$. It would clearly be of great importance to extend the stellar observations at lower 
metallicities to see whether the N/O ratio persists in its linear decrease or whether it flattens out and reaches values
similar to those of the ``low N/$\alpha$'' DLAs.

\subsection{N-Rich Stars}

Several definitions could be made of what is N-rich or N-normal, and some stars might be classified
differently according to the definition chosen. However, two stars in our sample that would surely be ascribed
to this group and are G64-12 and LP\,815-43. Neither of these two well-known metal-poor stars
shows any serious chemical anomaly. Both stars have a  Li abundance (Bonifacio \& Molaro 1997; 
Ryan, Norris \& Beers 1999; Bonifacio 2001) that places them on the lithium plateau.
This places serious constraints on the process that enriches them in N, since Li is destroyed at the temperatures
at which the CNO cycle is operating Li. The problem is not new and several other
N-rich field stars are known (Bessell \& Norris 1982; Laird 1985), all of which are lithium-normal 
(Spite \& Spite 1986). Likewise, the globular cluster NGC\,6397 is N-rich although Li-normal (Bonifacio et al.\ 2002).
In their pioneering survey Carbon et al.\ (1987) pointed out that the number of N-rich stars is very limited; 
however, in our example there are at least two clear examples which are similar to the previously five known (HD\,25329,
HD\,74000, HD\,97916, HD\,160617 and HD\,166913). We believe these stars are the exception rather than the rule, 
although they are perhaps less rare than previously thought.

\subsection{Nitrogen Nucleosynthesis and Final Remarks}

Nitrogen is one of the most abundant elements in the Universe. 
With C and O, it shares an important 
nucleosynthetic role as a key ingredient in hydrogen-burning
through the CNO cycle. However while for carbon and oxygen
the dominant production modes are the $\alpha$-chain reactions,
which start from the carbon-producing triple-$\alpha$ reaction,
for nitrogen the dominant production mode lies in the re-arrangement
of nuclei which occurs in the CNO cycle.
In CNO-cycled material $^{14}$N is increased at the expense of 
$^{12}$C and $^{16}$O. However, there is still considerable uncertainty 
as to where and when this process takes place.
Hydrogen-burning operates mainly through the CNO cycle
during the core H-burning (main sequence) phase of stars slightly
more massive than the Sun and during shell H-burning phases. The latter include:
the first ascent along the red giant branch (RGB), the horizontal
branch (HB) and  the asymptotic giant branch (AGB).
The first question to ask  is under which of these conditions the
nuclear-processed material is more readily shed into the
interstellar medium. Our targets are unevolved dwarfs and therefore the N and O
abundances discussed in this paper reflect conditions in the ISM at 
the time of star formation. The second question is where the C and O nuclei,
used to form N, come from. Were these synthesized by
a previous  stellar generation or by the star itself ?

In standard stellar models the material processed in the core of a massive star does not reach
the surface. Therefore, core H-burning should not provide any nitrogen. However, if
we add some extra mixing, e.g. because of rotation, a part of the core material may reach 
the surface (see the discussion in Meynet \& Maeder 2002 and their table 2).

Concerning shell-H burning, according to the standard models, it is only the AGB phase,
through the third dredge-up mechanism in which it is possible to bring nucleosynthesis products
to the surface. Also, in this case inclusion of rotation allows surface enrichment
even prior to the third dredge-up  (Meynet \& Maeder 2002).  

The second question we have posed is usually found in the literature as: Is nitrogen primary or
secondary? Since nitrogen needs C and O to be formed, if nitrogen is formed from
the C and O already existing at the time the star formed then it is called ``secondary''. If, instead,
the C and O are formed in the star itself and then used for N production then N is called ``primary''.
According to the standard models, primary production can take place in the AGB phase of stars that 
undergo hot bottom burning (Marigo 2001); therefore, it takes place in intermediate mass stars.
In models which include rotational mixing, production of primary nitrogen can instead take place 
both in intermediate mass and massive stars. Moreover, in intermediate mass stars the mechanism
may be different from that in the standard models: in an H-shell burning star the He-processed material,
containing C and O, may reach the H-burning shell through rotational diffusion, thus giving rise 
to primary nitrogen.

It thus appears reasonably clear, from the theoretical
point of view, that any stellar population will produce
some primary, as well as some secondary, nitrogen.
We thus expect this element to show a behavior that
is neither purely primary nor purely secondary.

Spectral syntheses presented in this article are subject to many uncertainties. Stellar parameters and line lists
are questioned first of all. While we did our best to obtain reliable stellar parameters and a well calibrated line list, 
two more uncertainties could still affect the final results: the effects of granulation/3D and the opacity uncertainties in 
the near UV. Both effects can be greatly minimized or perhaps even canceled if we measure 
the (N/O) ratios from the near-UV bands of NH and OH and plot them against [Fe/H] derived from the \ion{Fe}{ii} lines 
(Nissen et al.\ 2002). 
However, these effects may not be masked on the graphs which plot (N/0) against (O/H) (or (N/H)) since the latter
will be affected. While it is still impossible to carry out NH analysis with 3D models in a large number of metal-poor 
stars, we can guess what the general effect would be. The 3D effect on OH increases with decreasing [Fe/H] 
(Asplund \& Garc\'{\i}a P\'erez 2001). Thus, the lowest metallicity points on Fig.~\ref{fig5} will be shifted horizontally 
to the right and the signature of primary Nitrogen found in 1D models (i.e. the flattening at low (O/H)) will 
disappear. Finally, we note that even this ``guess'' is very uncertain since 3D models still do not resolve the
oxygen conflict in metal-poor stars, and more work is needed to achieve consistent abundance analysis.

\begin{acknowledgements} We would like to thank Andre Maeder and Georges Meynet for many important discussions and
comments on this paper. The referee Richard Henry is thanked for many valueable suggestions and comments.  
\end{acknowledgements}

%---------------------------bibliography---------------------------

\bibliographystyle{aa}

\listofobjects

\end{document}